\documentclass[prd,reprint,aps,tightenlines,preprintnumbers,amsmath,amssymb,
showpacs,superscriptaddress]{revtex4-1}
\usepackage{amsmath,amssymb,graphicx,natbib}
\usepackage{graphicx}
\usepackage{dcolumn}
\usepackage{bm}
\usepackage[hyperindex,breaklinks]{hyperref}
\usepackage{amsmath}
\usepackage{array}

\def\be{\begin{equation}}
\def\ee{\end{equation}}
\def\ba{\begin{eqnarray}}
\def\ea{\end{eqnarray}}
\def\ge{\mathrel{\raise.3ex\hbox{$>$\kern-.75em\lower1ex\hbox{$\sim$}}}}
\def\la{\mathrel{\raise.3ex\hbox{$<$\kern-.75em\lower1ex\hbox{$\sim$}}}}

\def\simgt{\mathrel{\raise.3ex\hbox{$>$\kern-.75em\lower1ex\hbox{$\sim$}}}}
\def\simlt{\mathrel{\raise.3ex\hbox{$<$\kern-.75em\lower1ex\hbox{$\sim$}}}}

\newcommand{\fr}[2]{\frac{#1}{#2}}

\newcommand{\nc}{\newcommand}

\nc{\gone}{\bar g_{\pi NN}^{(1)}}
\nc{\gzero}{\bar g_{\pi NN}^{(0)}}
\nc{\al}{\alpha}
\nc{\ga}{\gamma}
\nc{\de}{\delta}
\nc{\ep}{\epsilon}
\nc{\ze}{\zeta}
\nc{\et}{\eta}

\nc{\Th}{\Theta}
\nc{\ka}{\kappa}
\nc{\rh}{\rho}
\nc{\si}{\sigma}
\nc{\ta}{\tau}
\nc{\up}{\upsilon}
\nc{\ph}{\phi}
\nc{\ch}{\chi}
\nc{\ps}{\psi}
\nc{\om}{\omega}
\nc{\Ga}{\Gamma}
\nc{\De}{\Delta}
\nc{\La}{\Lambda}
\nc{\Si}{\Sigma}
\nc{\Up}{\Upsilon}
\nc{\Ph}{\Phi}
\nc{\Ps}{\Psi}
\nc{\Om}{\Omega}
\nc{\ptl}{\partial}
\nc{\del}{\nabla}
\nc{\ov}{\overline}
\nc{\newcaption}[1]{\centerline{\parbox{15cm}{\caption{#1}}}}

\newcommand{\deu}{D}

\newcommand{\qua}{$^4$He}

\newcommand{\sep}{$^{7}$Li}

\def\beq{\begin{equation}}
\def\eeq{\end{equation}}
\def\bmat{\begin{displaymath}}
\def\emat{\end{displaymath}}
\def\bear{\begin{eqnarray}}
\def\eear{\end{eqnarray}}
\def\bery{\begin{array}}
\def\ery{\end{array}}
\def\bit{\begin{itemize}}
\def\eit{\end{itemize}}
\def\ben{\begin{enumerate}}
\def\een{\end{enumerate}}
\def\btab{\begin{tabular}}
\def\etab{\end{tabular}}
\def\btbl{\begin{table}}
\def\etbl{\end{table}}
\def\bfig{\begin{figure}[htb]}
\def\efig{\end{figure}}
\def\bpic{\begin{picture}}
\def\epic{\end{picture}}

\def\ga{\mathrel{\raise.3ex\hbox{$>$\kern-.75em\lower1ex\hbox{$\sim$}}}}
\def\la{\mathrel{\raise.3ex\hbox{$<$\kern-.75em\lower1ex\hbox{$\sim$}}}}
\def\gappeq{\mathrel{\rlap {\raise.5ex\hbox{$>$}}
{\lower.5ex\hbox{$\sim$}}}}
\def\lappeq{\mathrel{\rlap{\raise.5ex\hbox{$<$}}
{\lower.5ex\hbox{$\sim$}}}}

\def\gyr{{\rm \, G\kern-0.125em yr}}
\def\mev{{\rm \, Me\kern-0.125em V}}
\def\gev{{\rm \, Ge\kern-0.125em V}}
\def\tev{{\rm \, Te\kern-0.125em V}}

\begin{document}

\title{Modified big bang nucleosynthesis with non-standard neutron sources }

\author{Alain Coc}
\affiliation{Centre de Sciences Nucl\'eaires et de Sciences de la
Mati\`ere  (CSNSM),
IN2P3-CNRS and Universit\'e Paris Sud 11, UMR 8609, B\^at. 104, 91405 Orsay Campus, France}

\author{Maxim~Pospelov}
\affiliation{Department of Physics and Astronomy, University of Victoria, 
Victoria, BC V8P 5C2, Canada}
\affiliation{Perimeter Institute for Theoretical Physics, Waterloo, ON N2J 2W9, 
Canada}

\author{Jean-Philippe Uzan}
\affiliation{Institut d'Astrophysique de Paris, Universit\'e Pierre \& Marie Curie - Paris VI,
CNRS-UMR 7095, 98 bis, Bd Arago, 75014 Paris, France}
\affiliation{Sorbonne Universit\'es, Institut Lagrange de Paris, 98 bis bd Arago, 75014 Paris, France}

\author{Elisabeth Vangioni}
\affiliation{Institut d'Astrophysique de Paris, Universit\'e Pierre \& Marie Curie - Paris VI,
CNRS-UMR 7095, 98 bis, Bd Arago, 75014 Paris, France}
\affiliation{Sorbonne Universit\'es, Institut Lagrange de Paris, 98 bis bd Arago, 75014 Paris, France}

\begin{abstract}
During big bang nucleosynthesis, any injection of extra neutrons around the time of the 
$^7$Be formation, i.e. at a temperature of order $T \simeq 50$~keV, can reduce the predicted 
freeze-out amount of $^7$Be + $^7$Li that otherwise remains in sharp contradiction with the 
Spite plateau value inferred from the observations of Pop II stars. However, the growing confidence in the 
primordial D/H determinations puts a strong constraint on any such scenario. We adress this issue in detail, 
analyzing different temporal patterns of neutron injection, such as decay, annihilation, resonant annihilation, and oscillation between mirror and standard model world neutrons. For this latter case, we derive the realistic injection pattern taking into 
account  thermal effects (damping and refraction) in the primordial plasma. If the extra neutron supply is the 
sole non-standard mechanism operating during the BBN, the suppression of lithium abundance below Li/H~$\leq 1.9 \times 10^{-10}$ 
always leads to the overproduction of deuterium, D/H~$\geq 3.6 \times 10^{-5}$, well outside the error bars suggested by recent observations. 

\end{abstract}

\maketitle

\section{Introduction}\label{sec1}

The  $\Lambda$CDM model of cosmology continues to withstand 
all observational tests of modern precision cosmology, and its triumph can only be compared to the 
similarly impressive performance of the Standard Model (SM) of particles and fields. 
Among the most nontrivial tests of the standard cosmological paradigm is the comparison of the 
Big Bang Nucleosynethesis (BBN) predictions, ever sharpened by the independent CMB-based determination of the 
baryon-to-photon ratio $\eta$, with observations. The latest most precise determination is from the 
Planck collaboration, $\eta = 6.047 \pm 0.074$ \cite{Ade:2013zuv}.

BBN respresents an early cosmological epoch ($t\simeq 200$ s), when the process of expansion and 
cooling of the Universe resulted in the creation of a few stable nuclei besides hydrogen. Its main effect is the 
creation of the sizable amount of helium. The determination of the helium abundance and its extrapolation to the 
primordial value is in perfect agreement with BBN predictions, once all sources of systematic errors are taken 
into account (see {\em e.g.} current review \cite{PDG} and references therein). 
Besides $^4$He, the BBN produces other light elements, and of particular interest for cosmology is the 
amount of primordial deuterium, surviving from incomplete burning at the BBN times. The determination of primordial 
deuterium abundance is a thorny issue in cosmology, as observations are difficult and performed only in a handful of 
damped Lyman-$\alpha$ systems. For a while, the scatter between different observations was significantly larger than the error bars would imply,
which could have been an indication for the deuterium depletion. However, over the course of the last two years, significant advances 
in the determination of D/H have been made \cite{oai:arXiv.org:1205.3785}, and the recently re-analyzed data point to a remarkable result \cite{Cooke:2013cba}
\be
{\rm D/H }= (2.53 \pm 0.04)\times 10^{-5}.
\label{cooke}
\ee
This result is in good agreement with the BBN predictions, see {\em e.g.} recent evaluations in Ref.~\cite{Coc:2013eea}, and  
has strong implications for many non-standard modifications of the cosmological model. 

Unlike deuterium, another trace element, $^7$Li, has been ``problematic" for over a decade. (For a detailed exposition of the problem,
see {\em e.g.} the dedicated reviews \cite{oai:arXiv.org:1203.3551,SpiteSpite12}.) The problem stems from the discrepancy of the 
BBN prediction with the primordial value for $^7$Li/H extracted from the absorption spectra in the atmospheres of the old stars. The absence of 
scatter in $^7$Li/H, and its remarkable constancy as a function of metallicity was discovered more than thirty years ago by F.~Spite and M.~Spite
\cite{Spite}. Throughout the 90-s, the Spite plateau value was believed to be a fair representation of the primordial value, and was widely used 
for the extraction of $\eta$. At the current value for $\eta$, it is well-known that the dominant fraction of predicted $^7$Li comes 
initially in the form of $^7$Be, which later on undergoes the capture process and becomes $^7$Li.  Current BBN predictions~ \cite{Coc:2013eea},
\begin{equation}
^7{\rm Li/H}_{\rm BBN}= (4.89^{+0.41}_{-0.39})\times 10^{-10},
\end{equation}
are a factor of $\sim 3-5$ larger than the Spite plateau value, $(1.23^{+0.34}_{-0.16})\times10^{-10}$ 
\cite{Ryan}, and  $(1.58\pm{0.31})\times10^{-10}$ 
\cite{Sbordone}, and many $\sigma$ away from it.\\

The goal of our paper is two-fold. Firstly, we would like to update the details of the neutron injection mechanism in one particular model
based on neutron-mirror-neutron oscillation. Earlier work by three of us on the subject \cite{Coc:2013eha} has to be extended to include the
thermal modification of the oscillation effects that will affect both the strength and the temporal pattern of the neutron injection
due do the oscillation from the mirror world. It is often the case that the 
injection of extra neutrons in models with decaying or annihilating particles 
is accompanied by additional non-thermal effects, and in that sence nBBN with mirror matter is the
``cleanest" realization of extra-neutrons scenario, as non-thermal effects are absent. 
 The second goal of our paper is to scan over the temporal patterns of the neutron 
injection of various types to determine whether this mechanism {\em by itself} is a sufficient reducer of $^7$Li/H that can also keep 
deuterium abundance consistent with observations. This second part can be viewed as an extension of the previous studies 
\cite{oai:arXiv.org:astro-ph/0402344,oai:arXiv.org:0906.2087,oai:arXiv.org:1208.0443}.\\

This paper is organized as follows. After discussing the different possible solutions to the lithium problem in Section~\ref{sec2}, Section~\ref{sec3} details the realistic pattern for the $n-n'$ oscillations in the presence of mirror matter taking into account thermal effects. In section~\ref{sec4} we compare different temporal patterns of neutron injection to  find out if any nBBN scenarios are consistent with both $^7$Li and D abundances. We reach our conclusion in Section~\ref{sec5}. 

\section{Possible solutions to the lithium problem}\label{sec2}

At this point, it is entirely not 
clear what resolves the lithium problem, and several logical pathways towards the resolution have been pursued (see e.g. Ref.~\cite{LiinC}):

\begin{enumerate}
\item    The amount of predicted $^7$Li is more sensitive than $^4$He to the adopted values for the nuclear reaction rates. 
While the main reactions determining the abundance of $^7$Li are now known with sufficient accuracy, for a while 
there was a possibility that some subdominant channels could increase the burning of $^7$Be \cite{oai:arXiv.org:astro-ph/0309480}. 
After much scrutiny \cite{oai:arXiv.org:astro-ph/0508454}, such possibilities look increasingly unlikely. 

\item  The stars are known to deplete heavier elements from their photosphere. The atomic diffusion at the bottom of the 
convective envelope (finely counterbalanced by the turbulent mixing) 
is often invoked as a possible mechanism for depleting lithium in Pop II stars \cite{oai:arXiv.org:astro-ph/0608201} . While certain amount of depletion will indeed 
happen for all stars, it is far from clear that it can occur uniformly for all stars along the Spite plateau without destroying its 
uniformity. In recent years, further questions are raised by the discovery of the ``meltdown" of the Spite plateau for the metallicities 
below $-3$ \cite{Sbordone}, for which no convincing explanation is found so far. 

\item  It is important to keep in mind that all lithium observations are made within stars that were born within or accreted to the Milky Way Galaxy
and its satellites, while the determination of $\eta$ is global. One cannot exclude some rather exceptional cosmological models where the 
uniformity of matter distribution is sacrificed and {\em e.g.} local value for $\eta$ is a factor of 3 lower than globally, leading to an ``accidental" local lithium underabundance \cite{oai:arXiv.org:0907.3919,regis}. 

\item Finally, particle physics may come to rescue and provide a modification to the standard BBN scenario in such a way that
 the lithium abundance is modified. Among most promising pathways are models with hadronic energy injection at the time of the 
BBN, or catalysis of certain nuclear reactions by the presence of negatively charged relics. For a review of possible options see 
{\em e.g.} \cite{oai:arXiv.org:0906.2087}.

\end{enumerate}

To summarize this discussion: because of inherent doubts about the fidelity with which the Spite plateau reproduces the primordial lithium abundance, 
it is admissible to think that the cosmological lithium problem may indeed be in a category of the ``astrophysical puzzles" 
rather than be an immediate make-or-break challenge to the standard cosmological paradigm. 

In this paper we give a further look into a problem of non-standard BBN with additional neutron injection (nBBN)
by a beyond-SM source.  It was recognized by Reno and Seckel in the 1980s that this class of scenarios will lead to the suppression of the 
freeze-out abundance of $^7$Be \cite{RenoSeckel}. This mechanism works by enhancing the conversion of beryllium to lithium, $^7$Be($n,p$)$^7$Li,
immediately after $^7$Be is created,
followed by more efficient proton burning of $^7$Li,  $^7$Li$(p,\alpha$)$\alpha$. After the CMB-based determination of $\eta$ and the emergence
of the cosmological lithium problem, this mechanism was further emphasized and investigated by Jedamzik \cite{oai:arXiv.org:astro-ph/0402344}, 
with many concrete particle physics realizations of the scenario built over the years \cite{oai:arXiv.org:hep-ph/0512044,oai:arXiv.org:1006.4172}. 
It is also well known \cite{oai:arXiv.org:astro-ph/0402344,oai:arXiv.org:0906.2087,oai:arXiv.org:1203.5701} 
that nBBN will cause a rise in the abundance of D/H, and 
given new tight constraints, (\ref{cooke}), one may question if the neutron injection 
mechanism is still a valid agent for reducing the cosmological abundance of lithium. 

\section{Neutron-mirror-Neutron oscillation in the early Universe}\label{sec3}

\subsection{Mirror matter models}

A mirror sector is constructed by assuming that the gauge group $G$ of the matter sector  is doubled to the product $G\times G'$.  Imposing a mirror parity under the exchange $G\leftrightarrow G'$ implies that the Lagrangian of the two sectors, ordinary and mirror, are identical so that they have the same particles content such that ordinary (resp. mirror) matter fields belonging to $G$ (resp. $G'$) are singlets of $G'$ (resp. $G$). They also have the same fundamental constants (gauge and Yukawa couplings, Higgs vev). The latter point implies that the microphysics (and in particular the nuclear sector) is identical in both sectors. The two sectors are coupled through gravity, and can eventually interact via some couplings so that the general form of the matter Lagrangian is
$$
{\cal L} = {\cal L}_G(e,u,d,\phi,\ldots) + {\cal L}_G(e',u',d',\phi',\ldots) +{\cal L}_{\rm mix}.
$$
Such a sector was initially proposed by Li and Yang~\cite{LiYang} in an attempt to restore global parity symmetry and was then widely investigated~\cite{mirror-gen,sterile}. Any {\em neutral} ordinary particle, fundamental or composite, can be coupled to its mirror partner hence leading to the possibility of oscillation between ordinary and mirror particles.  For instance a mixing term of the form ${\cal L}_{\rm mix}\propto F_{\mu\nu}'F^{\mu\nu}$ will induce a photon-mirror photon oscillation, ordinary neutrinos can mix with mirror neutrinos and oscillate in sterile neutrinos~\cite{sterile}. Among all the possible mixing terms, special attention has been drawn~\cite{mirrorneutron} to the mixing induced between neutrons and mirror neutrons. Such a possibility is open as soon as ${\cal L}_{\rm mix}$ contains a term $\propto (udd)(u'd'd') + (qqd)(q'q'd')$; see e.g. Ref.\cite{mirrorneutron} for details. It was also pointed out~\cite{mirrorneutron}  that a neutron--mirror neutron oscillation could be considerably faster than neutron decay, which would have interesting experimental and astrophysical implications.

\subsection{$n-n'$ oscillations}

We begin by analyzing $T=0$ case for the oscillation between ``our world" neutron $n$
and the ``mirror world" neutron-like particle $n'$. We will assume an {\em approximate} mirror symmetry that sets 
the masses of $n$ and $n'$ particles nearly equal, so that in 
$m_{n'} = m_{n} +\Delta m$ relation,
$\Delta m \ll m_{n,n'}$. We will allow for the interaction between the two sectors, 
that mixes the wave functions of normal and mirror neutrons, 
\begin{eqnarray}
{\cal H } = (\bar n \bar{n}'){\cal M}
 \left(
\begin{array}{c}
n\\ n' 
\end{array}
\right); ~~{\cal M} =
\left(
\begin{array}{cc}
\Delta m-\frac{i}{2}\Gamma_n& m_{12} \\ m_{12}^*& -\frac{i}{2}\Gamma_{n'}
\end{array}
\right).
\end{eqnarray}
$\Gamma_{n,n'}$ are the decay rates of $n,n'$.
Without loss of generality one can take the mixing parameter $m_{12}$ in the mixing matrix ${\cal M}$ 
to be real and positive. There is a significant freedom in the choice of the parameters $\Delta m$ and $m_{12}$, 
limited only by the experiments with ultracold neutrons, and by theoretical 
considerations related to the compositeness of $n$ and $n'$. The quark composition of $n$, and presumably a similar 
quark$'$ composition of $n'$ dictates that $m_{12}$ parameter is not ``elementary", but in fact is a descendant of a 
higher-dimensional operator that connect normal and mirror sectors. The lowest dimension 6-quark operator responsible for such 
mixing will be given by 
\be
{\cal L} _{\rm mix} = \frac{1}{\Lambda^5} \bar {\eta}_n \eta_{n'} +(h.c.)
\ee
where $\Lambda$ is roughly the high-energy scale where such operator is generated, and $\eta_n$ and $\eta_{n'}$
are the three-quark currents that interpolate between vacuum and $n$ states: 
$\eta_n = 2\epsilon_{abc} (d_a^T C\gamma_5 u_b) d_c$ with an analogous expression for $n'$. 
It is fair to take $\Lambda$ at the weak 
scale and above (given no signs of new physics at the LHC), $\Lambda \geq 300$ GeV. The matrix element of the 
$\eta_n$ current is known from hadronic physics, $\langle 0| \eta_n|n\rangle \simeq n(x)\times  0.02 {\rm GeV}^3 $.
Taking same matrix element in the mirror sector, we arrive at the following matching condition, 
\be
m_{12} = 4\times 10^{-4} ~ {\rm GeV}^6 \times \Lambda^{-5} \Longrightarrow m_{12} \la 2\times 10^{-7}
{\rm eV}. 
\label{limit12}
\ee
We conclude that mixing matrix elements below $10^{-7}$ eV are in general compatible with the composite nature of 
nucleons and the absence of new physics below the weak scale. 

\subsection{Experimental constraints on ($\Delta m,m_{12}$)}

We next address the question of what experimental constraints on the combination of $\Delta m$ and $m_{12}$ the precision measurements 
with neutrons would impose. Interestingly this issue had seen some lively debates, and is not as straightforward as it may sound. 
Starting from the mass matrix ${\cal M}$, one can derive the zero-temperature probability for the $n\leftrightarrow n'$ oscillation,
\be
\left.P_{n\leftrightarrow n'} \right|_{T=0} = \frac{(2m_{12})^2\sin^2\left[\frac12t\sqrt{\Delta m ^2 + (2m_{12})^2}\right]}{\Delta m ^2 + (2m_{12})^2}\times {\rm e}^{-\Gamma_nt},
\label{T=0}
\ee
where we have also set $\Gamma_n = \Gamma_{n'}$. The combination $(2m_{12})^2(\Delta m ^2 + (2m_{12})^2)^{-1}$ is often called 
$\sin^2(2\theta)$. 
In the limit of exact mirror symmetry, $\Delta m = 0$, this formula corresponds to the $n\leftrightarrow n'$  oscillation probability
with the maximal $\theta= \pi/4$ mixing.
Experimental constraints on $P_{n\leftrightarrow n'}$ can be derived from the analysis of the neutron life-time experiments
\cite{Mampe:1989xx}. For example, the analysis performed in Ref.~\cite{Serebrov:2007gw} quotes the limit on $m_{12}$ under strict 
mirror symmetry $\Delta m=0$, $m_{12} < 1.5\times 10^{-18}$~eV. 
The point of contention in these limits is often in an extra assumption 
of no extra contributions to 11 and 22 elements of ${\cal M}$ from the magnetic fields that an experimenter can control 
and mirror magnetic field (that is beyond his/her control) \cite{Berezhiani:2008bc,Berezhiani:2012rq}. 

In what follows we are going to consider the following hierarchical pattern, 
\be
\label{hierarchy}
\tau_n^{-1} \ll m_{12} \ll \Delta m \ll 10^{-7}~{\rm eV},
\ee
where $\tau_n$ is the neutron lifetime. To satisfy experimental constraints on oscillations, we are going to 
adopt the limit on time-average of $P_{n\leftrightarrow n'}$ obtained in Ref.~\cite{Sarrazin:2012sc}, $P_{n\leftrightarrow n'}  < 7\times 10^{-6}$,
which is derived without assuming $\Delta m= 0$. For the chosen hierarchy (\ref{hierarchy}) this limit implies
\be
\label{explimit}
\frac{m_{12}^2}{\Delta m^2} \la 3\times 10^{-6}.
\ee
Notice  that once (\ref{limit12}) and (\ref{explimit}) are satisfied, in principle both $m_{12}$ and $ \Delta m$ can be much larger 
than the inverse of the neutron lifetime in vacuum, and much larger than the Hubble rate during the BBN, 
\be
H = \frac{1}{2t} \simeq \frac{T_9^2}{356~{\rm s}}.
\ee
Here $T_9$ is the photon temperature in units of $10^9$ K, and at the BBN epoch relevant for $^7$Li+$^7$Be formation, 
$H$ is in the interval $\sim 10^{-3}-10^{-2}$Hz or $\sim10^{-18}-10^{-17}$ eV. 

\subsection{Effects on BBN}

The main point of this section is that under the conditions that exist in the early Universe, 
the oscillation probabilities are changed rather drastically. The physical reason for that is that the 
hypothesized $n\leftrightarrow n'$ oscillation is a quantum phenomenon that requires coherence 
in the phase of the wave function to be preserved. However, rapid rescaterings of neutrons on electrons and positrons, 
photons and protons (and presumably with similar processes in the mirror sector) leads to a rapid ``reset" of the quantum 
phase. The neutron collision rate $\Gamma_{\rm col}$ determines the coherence time interval, $\tau_{\rm coh} \sim 1/\Gamma_{\rm col}$, 
and in the regime when $\Gamma_{\rm col}$ is larger than any other dimensionful parameters, the 
time-average oscillation probability will scale as $P_{n\leftrightarrow n'}\propto m_{12}^2\Gamma_{\rm col}^{-2}$, and the 
{\em rate} for the neutron-mirror-neutron interconversion will be $\propto m_{12}^2\Gamma_{\rm col}^{-1}$.
These are very important modifications of the oscillation rate, and we address them below in a more quantitative manner. 

First, for the reasons explained in Ref. \cite{Coc:2013eha}, we assume that the temperature of the mirror world is smaller, as well as 
the number density of mirror baryons. This means that in the scattering processes the main contributions come from $n$ and not $n'$. 
Moreover, the decay rates for $n,n'$ particles are subdominant to the rescattering rates, and thus can be neglected in the calculation 
of the oscillation probability. We then have the 
following modification of the $n-n'$ mass matrix,
\be
\label{MofT}
{\cal M} \to {\cal M}_{\rm eff} = \left(
\begin{array}{cc}
\Delta m +\Delta m_{\rm eff} (T) -\frac{i}{2}\Gamma_{\rm eff}(T)& m_{12} \\ m_{12}& \approx 0
\end{array}
\right).
\ee
In this formula, the temperature-dependent mass shift $\Delta m_{\rm eff} (T)$ is induced by the real part of the neutron forward scattering amplitude, 
while the imaginary part, $\Gamma_{\rm eff}(T)$, by optical theorem is related to the total cross section. 
Since $m_{12}$ is very small, the process of $n\leftrightarrow n'$ oscillation is best described as the perturbation on top of the 
scattering processes that preserve number of neutrons. 
The oscillation rate is given by the rescattering rate mutiplied by the square of the {\em effective} mixing angle, and because of the thermal effects, 
$\theta_{\rm eff}\ll \theta$,
\begin{eqnarray}
\Gamma_{n\leftrightarrow n'} = \Gamma_{\rm eff}\times 
\left| \frac{m_{12}}{\Delta m +\Delta m_{\rm eff} (T) -\frac{i}{2}\Gamma_{\rm eff}(T) }  \right|^2\nonumber
\\
=\frac{(2m_{12})^2\Gamma_{\rm eff}(T)}{4(\Delta m +\Delta m_{\rm eff} (T))^2 +\Gamma^2_{\rm eff}(T) }.
\label{rate}
\end{eqnarray}
This treatment follows a well-established formalism for $K^0-\bar K^0$ oscillations that can be found
{\em e.g.}  in a textbook \cite{Okun:1982ap}. 

\begin{figure}
\begin{center}
\resizebox{0.9\columnwidth}{!}{\includegraphics{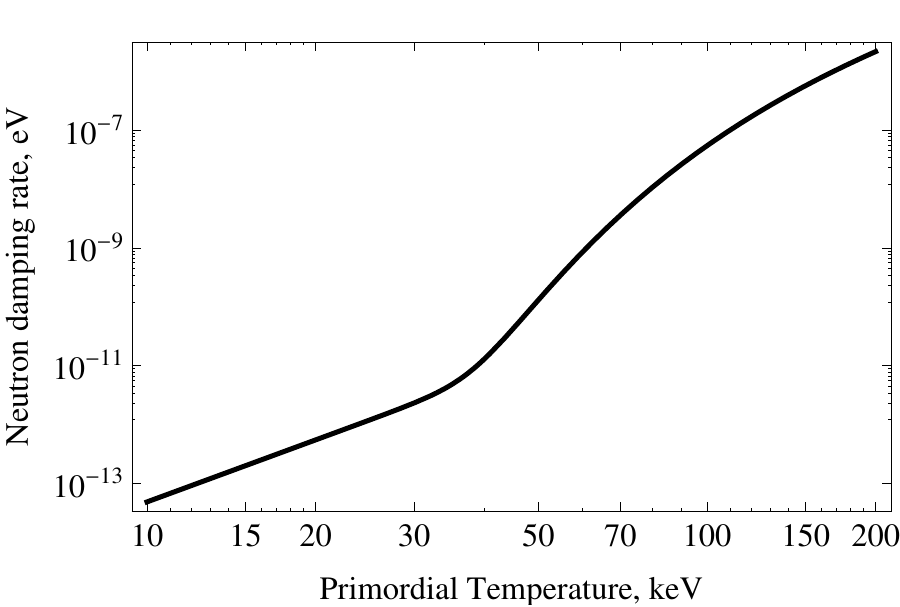}}
\caption{The neutron damping rate $\Gamma_{\rm eff}(T)$ is units of eV plotted as a function of temperature $T$, in units of keV.
The change from the predominantly electromagnetic to the strong force scattering occurs at $T\simeq 40$ keV, right after $^7$Be formation. 
}
\label{DampRate}
\end{center}
\end{figure}

According to general theory, the damping rate $\Gamma_{\rm eff}(T) $ can be expressed as
\be
\Gamma_{\rm eff}(T)  = \langle \sigma v  n \rangle,
\ee
where $v$ is the relative velocity between the neutron and scattering centers, and $n$ is their number density. The 
average is taken over the velocity distribution of particles in the bath. We will approximate  $\Gamma_{\rm eff}(T)$ by the sum of the 
two most important contributions: electromagnetic scattering on electrons and positrons and strong force scattering on protons. Direct calculation gives
\begin{eqnarray}
\label{Gamma}
\Gamma_{\rm eff}(T) = \sigma_{np}\times 4\sqrt{\frac{T}{m_p \pi}}\times n_p + ~~~~~~~~~~~~~~~~\\ \nonumber
\frac{2 \pi \alpha^2\mu_n^2}{m_p^2}\left(\fr12 +\log \left[ \frac{2(2m_eT)^{1/2}}{\omega_p} \right]\right)2\sqrt{\frac{2T}{m_e \pi}}\times(n_e+n_{\bar e}).
\end{eqnarray} 
In this expression, $\sigma_{np} \simeq 20$ bn is the low-energy cross section for $n-p$ scattering, 
$\mu_n \simeq -1.9$ is the neutron's magnetic moment in units of nuclear magneton, 
$n_p$ is the number density of protons (cross sections on $^4$He is much smaller and helium contribution can be neglected),
and $n_e+n_{\bar e}$ is the exponentially diminishing number density of electron positron pairs,
\be
n_e+n_{\bar e} \simeq \sqrt{\fr{2}{\pi^3}}(m_e T)^{3/2}\times \exp[ -m_e/T]  
\ee
This number density also defines the plasma frequency that enters the Coulomb logarithm in Eq.~(\ref{Gamma}),
$\omega_p^2 = 4\pi\alpha/m_e \times (n_e+n_{\bar e} )$. The plot of $\Gamma_{\rm eff}(T)$ is shown in Fig. 1. 
As one can see, there is a kink in $\Gamma_{\rm eff}(T)$ at $T\simeq 40$~keV, signalling the change from the 
scattering on electrons and positrons to the predominantly scattering on protons. Because of the relatively large value of 
$\Gamma_{\rm eff}(T)$ at early times, the oscillation between normal and mirror world neutrons will be  suppressed.

Next, we address the question of the effective mass shift $\Delta m_{\rm eff} (T)$ due to scattering. To that purpose, one needs to calculate the 
neutron forward scattering amplitude without change of the spin direction. 
Magnetic moment of the neutron does not contribute to the effect in the first order of perturbation theory, because 
it requires the spin flip. The scattering on protons then is the leading effect, and one can deduce that 
\be
\Delta m_{\rm eff} (T) \simeq -\fr{2\pi}{m_n}\times {\rm Re}\, f(0)\times n_p,
\ee
and ${\rm Re}\, f(0)$ can be taken directly from data on the $n-p$ scattering length. 
After working out the numerics, we conclude that the mass shift is not important
for the problem under consideration. It is true that since the $\Delta m$ sign is not known a priori, 
there is a possibility of a  cancellation between $\Delta m$ and $\Delta m_{\rm eff} (T) $ 
in the rate formula (\ref{rate}). However, the emergent resonance is not sharp, being dominate by 
$\Gamma_{\rm eff}$. This is the main reason why the mass shift effects can be neglected.

\begin{figure}
\begin{center}
\resizebox{0.9\columnwidth}{!}{\includegraphics{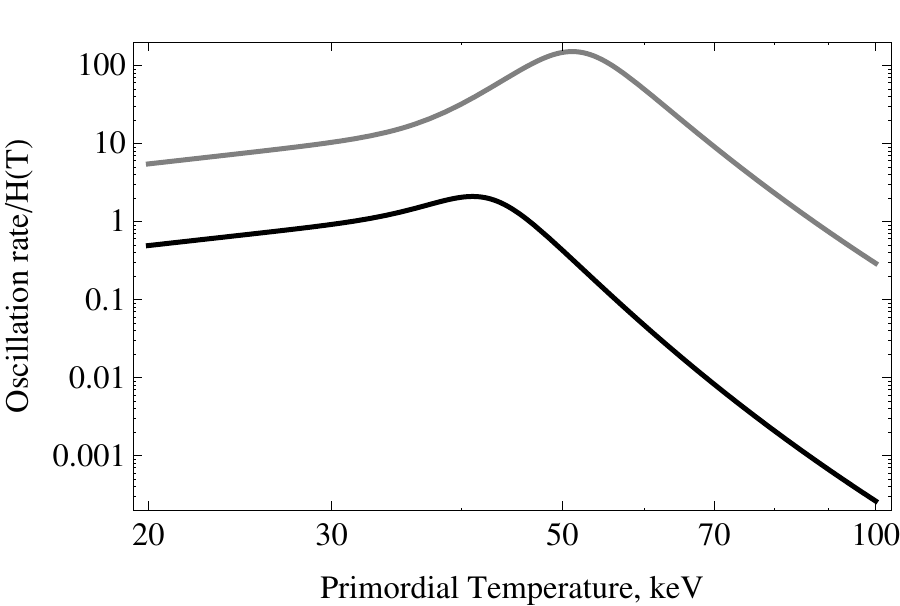}}
\caption{The $n\leftrightarrow n'$ oscillation rate normalized on the Hubble rate, 
$\Gamma_{n\leftrightarrow n'}/H$, as a function of temperature $T$, expressed in keV.  
 The top curve is for the choice  $\Delta m = 10^{-10}$~eV, and $m_{12} =10^{-13}$~eV,
and the bottom curve is for $\Delta m = 10^{-11}$~eV, and $m_{12} =3\times 10^{-15}$~eV.
The top curve becomes larger than one during $^4$He formation at $T\sim 80$~keV, while 
the bottom curve reaches one only for a brief period around $T\sim 40$~keV. }
\label{OscRate}
\end{center}
\end{figure}

Finally, we present several representative cases for the $n-n'$ oscillation rate in Fig. 2,
for different choices of $\Delta m$ and $m_{12}$. Of course, the most relevant parameter is the 
rate weighted by the Hubble expansion rate. When $\Gamma_{n\leftrightarrow n'} /H >1$, the oscillations are 
occuring efficiently, and if it is much smaller than one, the oscillation mechanism for changing neutron abundance 
 can be neglected. As Fig. 2 clearly demonstrates, the 
actual behavior is very sensitive to the underlying choice of $m_{12}$ and $\Delta m$. Only a sufficiently large value of 
$m_{12}$ can ensure $\Gamma_{n\leftrightarrow n'} /H >1$, and in particular the choice
of $\Delta m = 0$ and $m_{12} = 1.5\times 10^{-18}$~eV (border-line of the 
existing bounds in the exact mirror symmetry case), will lead to 
$\Gamma_{n\leftrightarrow n'} <10^{-5}$ at all times when the neutron-mediated $^7$Be burning is possible. 
Therefore, the only reasonable chance for reducing lithium abundance this way is to accept a small but non-zero value
for $\Delta m$.

\section{${\rm n}$BBN results}\label{sec4}

We now present our main results for nBBN focusing on 4 main mechanisms of neutron injection.

\subsection{Description of the models}

While the previous section describes in details the implementation of the oscillation of neutron with mirror world neutrons, there are three other possibilities  to inject neutrons during BBN. We thus consider the 4 models.

\begin{enumerate}
\item  {\em $n-n'$ oscillation}. This model has been described in the previous section and an early analysis was presented in Ref.~\cite{Coc:2013eha}. This model contains 2 physical parameters, $\Delta m$ and $m_{12}$ with $m_{12}/\Delta m <1$ and 3 cosmological parameters, $x$, the baryon-to-proton ratio in each world $\eta$ and $\eta'$. We shall assume that $\eta=\eta_{\rm CMB}$ and scan the other parameters.

\item  {\em Particle decay}. This class of models assumes the existence of an hypothetical particle $X$ that can decay and produce neutron. The decay rate $\Gamma$ is proportional to the abundance of the unstable particle and its liftime, $\Gamma \propto   (Y_X/ \tau_X) \exp(-t/\tau_X)$.
We scan over the initial abundance $Y_X$and the lifetime $\tau_X$, or equivalently $\lambda_0\sim Y_X/\tau_X$ so that we have 2 independent parameters to consider. 

\item  {\em Particle annihilation}. These models are characterized,  besides $Y_X$, by the annihilation rate. This channel is the slowest way for injecting neutrons.  It corresponds to the case  5 of Ref.~\cite{oai:arXiv.org:1208.0443}  with
a single parameter, $\lambda_0$. 

\item  {\em Resonant particle annihilation.} If a narrow resonance is present at some energy $E_r$, then the 
annihilation rate scales as $\exp(-E_r/T)$ \cite{Pospelov:2010hj}. In such a case the model depends onf the resonance energy, $E_r$, the 
abundance of annihilating particles, $Y_X$, and the annihilation strength, $\lambda_0$.
\end{enumerate} 

\begin{figure}[htb!]
\begin{center}
 \includegraphics[width=0.9\columnwidth]{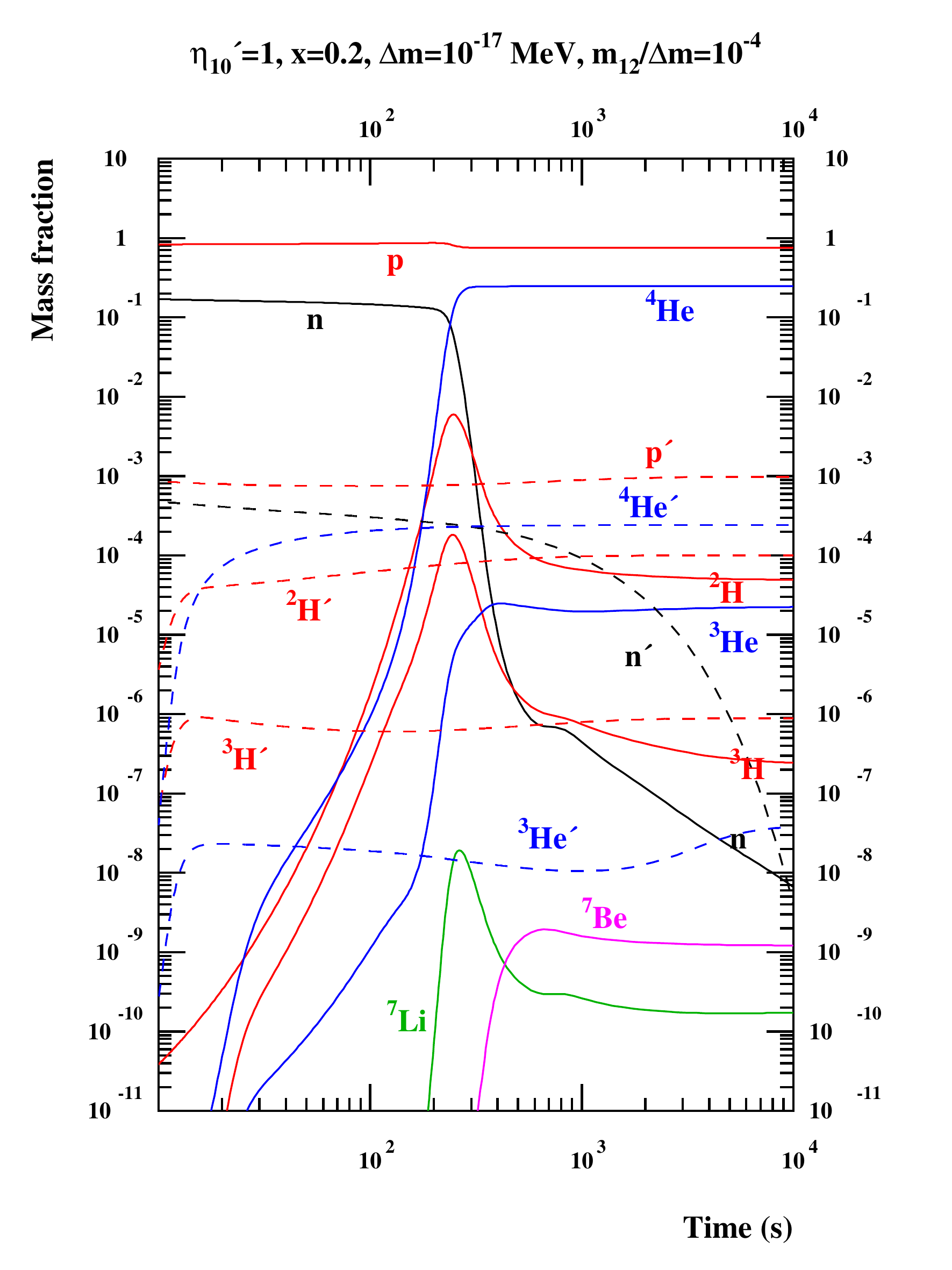}
\caption{\label{f:time} Time evolution of abundances in a model of $n-n'$ oscillation, assuming $\eta=\eta_{\rm CMB}$, $x=0.2$ and $\eta'=1$. 
The curves represent mass fractions of ordinary (solid) or mirror (dash) isotopes calculations, with only neutrons allowed to flow from one world to the other. Is shows, in our world, an increase of the neutron abundance resulting in a reduction of the Beryllium7 one. }
\end{center}
\end{figure}

These 4 classes of models allow one for a neutron injection during BBN, with different efficiencies. Table~\ref{tab1} summarizes the parameters on which they depend.

\begin{table}[htdp]
\caption{Summary of the 4 classes of models and of their free parameters (beside $\eta$).}
\begin{center}
\begin{scriptsize}
\begin{tabular}{l | c|c}
\hline\hline
Model & Physical parameters & Cosmological parameters \\
\hline
$n-n'$ oscillation & $\Delta m, m_{12}$& $x,\eta'$ \\
Particle decay & $\tau_X$ & $Y_X$ \\
Particle annihilation & $\lambda_0$ & $Y_X$ \\
Resonant annihilation &  $E_r$ & $Y_X$\\
\hline\hline
\end{tabular}
\end{scriptsize}
\end{center}
\label{tab1}
\end{table}%

\begin{figure*}[htb!]
\begin{center}
 \includegraphics[width=0.62\columnwidth]{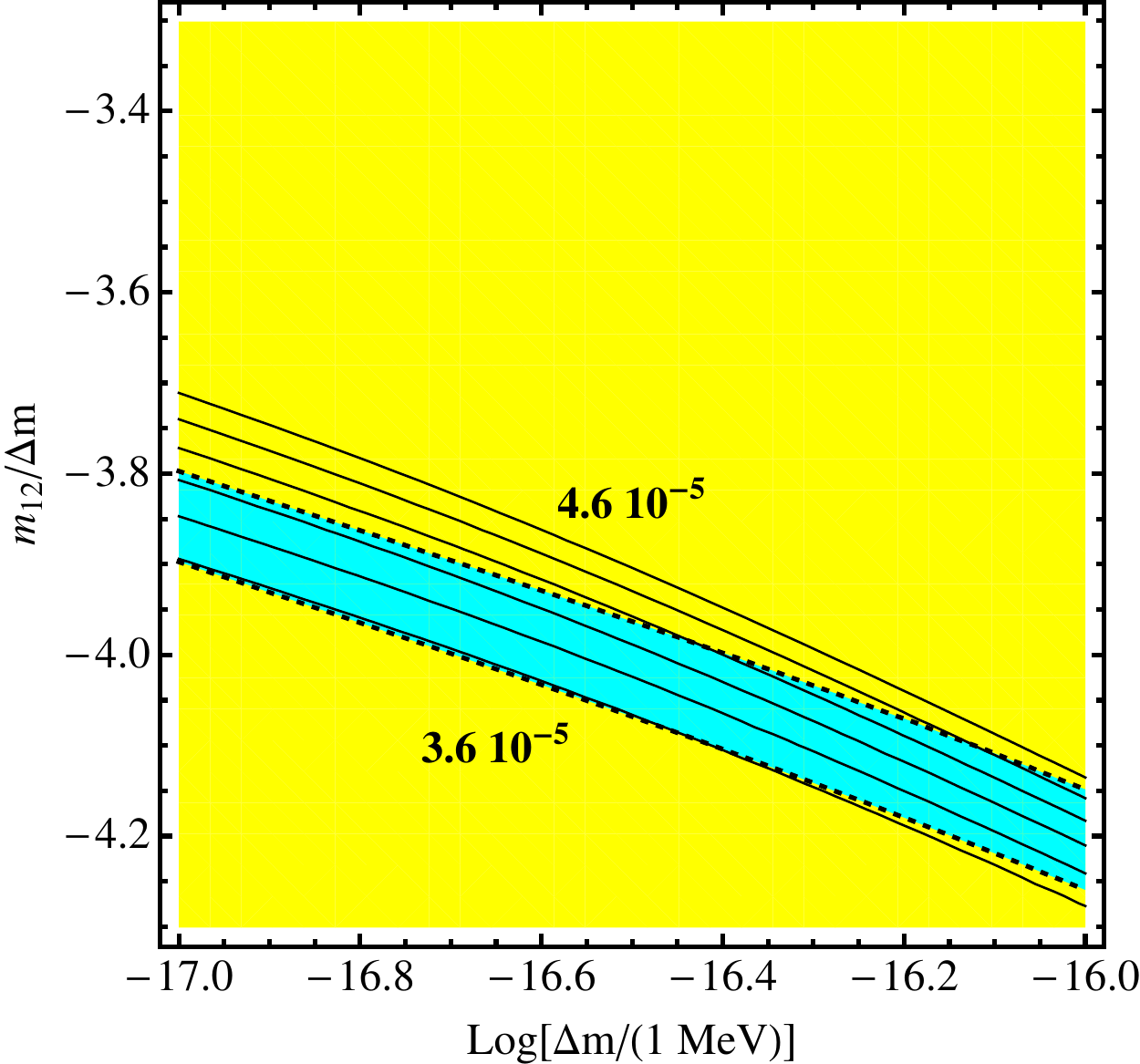} \includegraphics[width=0.62\columnwidth]{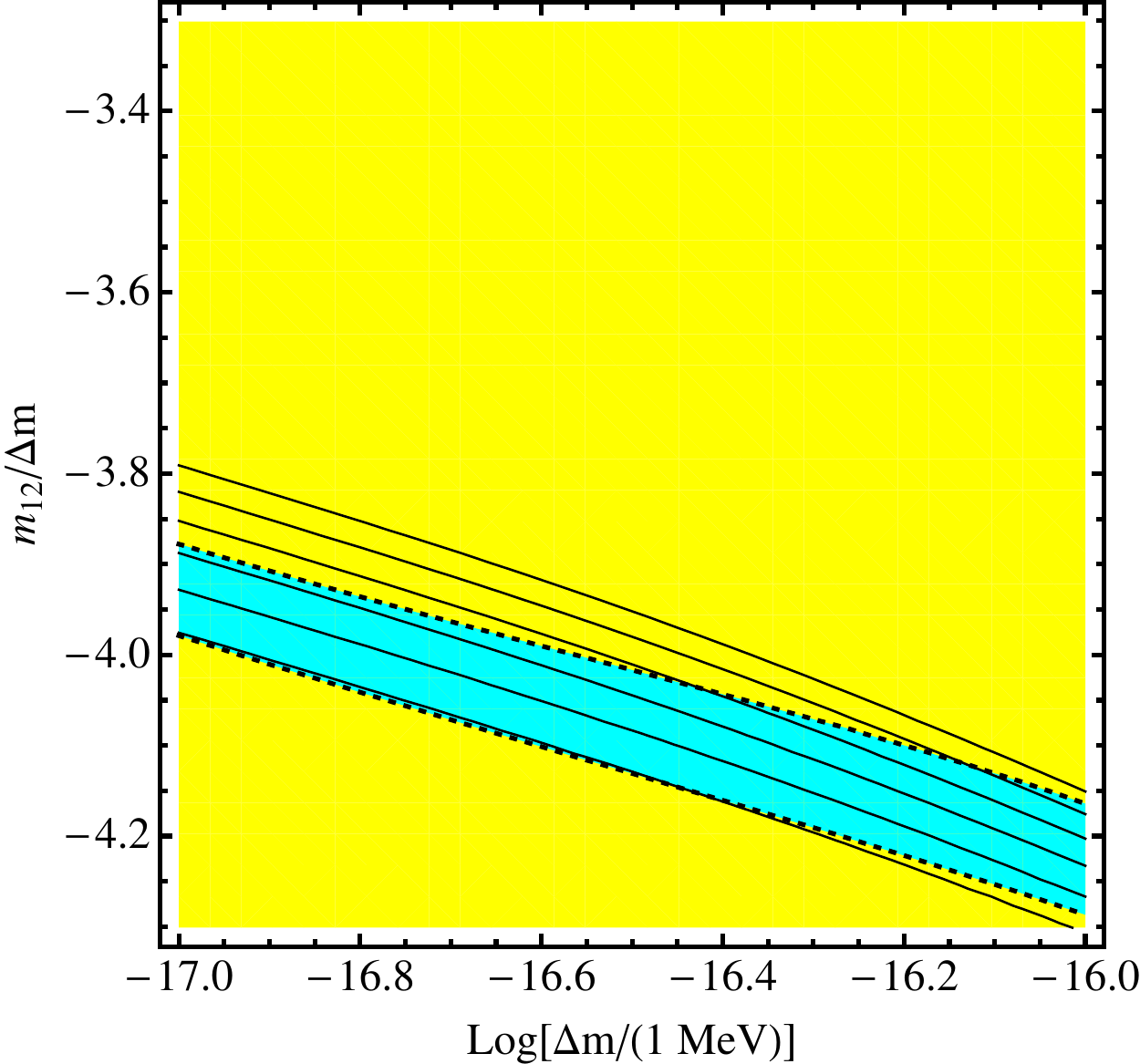} \includegraphics[width=0.62\columnwidth]{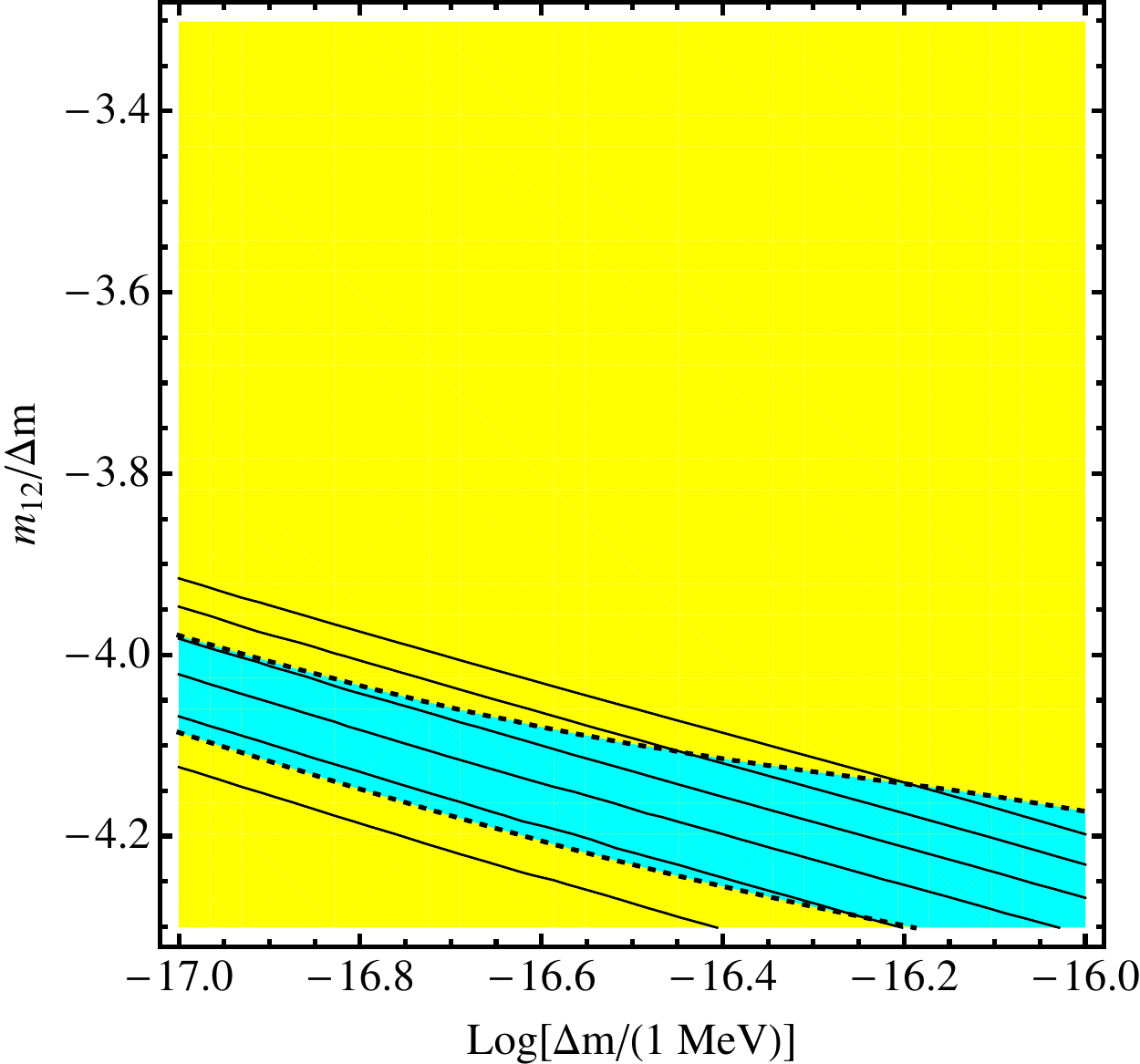}
\caption{ \label{f:osc1}$n-n'$ oscillation. Contour plots in the space of the two physical parameters $(\Delta m,m_{12})$ assuming $\eta=\eta_{\mathrm CMB}$ and $x=0.2$  respectively with $\eta'=10^{-10}$ (left) and $\eta'=3\times10^{-10}$ (middle) and $x=0.5$ and $\eta'=10^{-10}$ (right). The blue strip corresponds to models for which the BBN predictions are compatible with the observational constraints for both helium-4 and lithium-7. The solid  lines indicate the prediction of deuterium abundance D/H = \{3.6, 3.8, 4.0, 4.2, 4.4, 4.6\}$ \times10^{-5}$ from top to bottom.}
\end{center}
\end{figure*}

\subsection{Constraints from BBN}

In order to investigate if any of these models of neutron injection are compatible with light element abundance observations including lithium-7, we scan the parameters space of each model (see Table~\ref{tab1}) and display (in blue) the zone allowed by observations of \qua\ 
($0.2368 < Y_p < 0.2562$, yellow) and \sep\ ($1.27\times10^{-10}<{\rm Li/H} < 1.89\times10^{-10}$, blue) in figures 4,5. Indeed $\eta$ remains fixed to $\eta_{\rm CMB}$. Then, we should superpose the prediction of \deu\ observations ($2.49\times10^{-5}  < {\rm D/H} < 2.57 \times10^{-5}$). As we shall see, this zone would lie 
outside of the frame and we thus only display the 6 curves corresponding to D/H = \{3.6, 3.8, 4.0, 4.2, 4.4, 4.6\}$ \times10^{-5}$.

\vskip.5cm
\noindent{\em $n-n'$ oscillation}. We implemented the equations of Section~3 in our BBN code, which allows us to predict the evolution of the abundance of all light elements, in both the real and mirror worlds. Fig.~\ref{f:time} gives an example of the evolution of the different abundances as a function of time. It has to be compared to Figs. 6 and 7  of our previous work~\cite{Coc:2013eha} (where $^{2,3}$H' and $^3$He' were not displayed). It can be seen that the effect of the oscillation, from the standard world point of view, is an injection of neutron that modifies $n/p$ compared to standard BBN typically for $t>10^{3}$~s.

Figure~\ref{f:osc1} depicts the zone of the parameter space that allows one to reconcile the predicted lithium-7 abundance to its observed value, for different sets of the cosmological parameters in the mirror world. It is easily to conclude that forcing the model in such a way leads to a too large level of deuterium, typically larger than $3.6\times10^{-5}$ while observations require it to be of the order of $2.5\times10^{-5}$.

\begin{figure}[htb!]
\begin{center}
 \includegraphics[width=0.8\columnwidth]{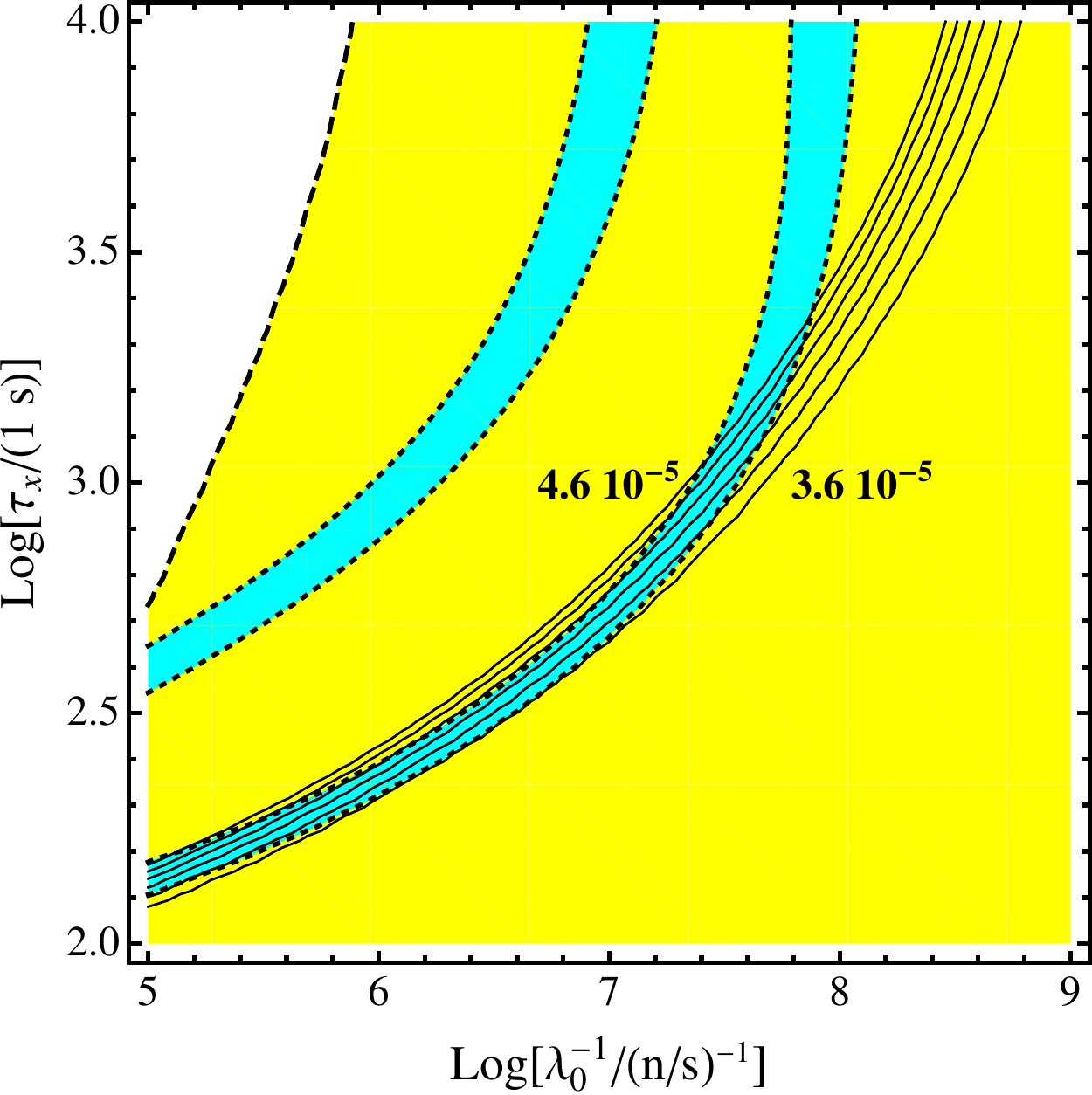}
\caption{\label{f:decay} {\it Decay of massive particles}. Contour plot assuming $\eta=\eta_{\rm CMB}$ for the two parameters of the model: the lifetime $\tau_x$ of the massive particle and the decay rate $\lambda_0\exp(-t/\tau_X)$. This can be compared to the case 4 of Ref.~\cite{oai:arXiv.org:1208.0443}. The solid dashed lines indicate the prediction of deuterium abundance D/H = \{3.6, 3.8, 4.0, 4.2, 4.4, 4.6\}$ \times10^{-5}$ from top to bottom.}
\end{center}
\end{figure}

\begin{figure}[htb!]
\begin{center}
 \includegraphics[width=0.8\columnwidth]{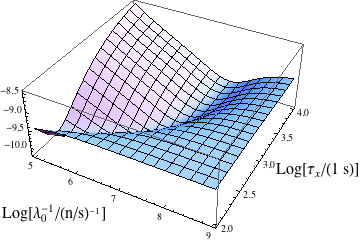}
\caption{\label{f:decay2} {\it Decay of massive particles}. The abundance of lithum-7 produced during BBN, as a function of the two parameters
 ($\tau_X,\lambda_0$) has a valley. See text for an explanation of the shape of this surface and compare with Fig.~\ref{f:decay}.}
\end{center}
\end{figure}

\vskip.5cm
\noindent{\em Particle decay}. We scan the parameter space ($\tau_X,\lambda_0$), keeping in mind that $Y_X\sim\lambda_0\tau_X$ and the result is depicted on Fig.~\ref{f:decay}. The morphology of the region compatible with helium-4 and lithium-7 (blue strip) is the result of the fact that the predicted shape of the surface \sep($\tau_X,\lambda_0$) has a valley (see Fig.~\ref{f:decay2}) that is intersected by the slab $1.27\times10^{-10}<{\rm Li/H} < 1.89\times10^{-10}$. 

\begin{table}[htdp]
\caption{Mass fractions of the different light elements produced during BBN for a model of particle decay (see Fig.~\ref{f:decay}) for different
values of the decay rate $\lambda_0$, assuming that $\tau_X=10^3$~s, quoted for $t=1.677\times 10^4$ s from the Big Bang.}
\begin{center}
\begin{scriptsize}
\begin{tabular}{|c|ccc|}
\hline
$\log\lambda_0^{-1}$ &    5.5    &  7  &     9 \\
\hline\hline
$n$ & $2.9\times10^{-7}$ &  $9.36\times10^{-9}$  & $1.62\times10^{-9}$ \\
$^{1}$H & $7.478\times10^{-1}$ & $7.535\times10^{-1}$& $7.537\times10^{-1}$\\
$^{2}$H & $5.578\times10^{-4}$ & $9.922\times10^{-5}$ & $4.131\times10^{-5}$ \\
$^{3}$H & $3.020\times10^{-6}$ & $4.775\times10^{-7}$ &  $2.029\times10^{-7}$\\
$^{3}$He & $5.577\times10^{-5}$ & $1.951\times10^{-5}$ & $2.353\times10^{-5}$  \\
$^{4}$He & $2.515\times10^{-1}$ & $2.463\times10^{-1}$ & $2.462\times10^{-1}$\\
$^{6}$Li & $8.940\times10^{-13}$ & $1.483\times10^{-13}$ & $6.143\times10^{-14}$ \\
$^{7}$Li &  $1.831\times10^{-9}$ & $3.116\times10^{-10}$  & $1.767\times10^{-10}$ \\
$^{7}$Be & $6.367\times10^{-13}$ & $4.939\times10^{-11}$ & $2.374\times10^{-9}$\\
\hline
\end{tabular}
\end{scriptsize}
\end{center}
\label{tab2}
\end{table}%

The limit $\log \lambda_0^{-1}\rightarrow+\infty$, or equivalently $\lambda_0\rightarrow0$, corresponds to the standard BBN limit. This explains why the right part of the parameter space is compatible with helium-4. The smaller $\log \lambda_0^{-1}$ the higher is the neutron injection so that in the left part of the plot, BBN overproduces both lithium-7 and helium-4. As can be concluded from Table~\ref{tab2}, at high $\log\lambda_0^{-1}$ the neutron injection is too small so that the destruction of $^7$Be due to neutron capture remains too small. This corresponds to an almost standard BBN. When $\log\lambda_0^{-1}$ decreases, the neutron production increases which allows to reduce $^7$Be enough for the final lithium-7 abundance to be reconciled with observation. This corresponds to the right blue strip which is dominated by the channel $^4{\rm He}+ ^3{\rm He}\rightarrow  ^7{\rm Be}+\gamma$ followed by a $\beta$-decay. Between the two blue strips the final abundances of Lithium-7 is too low. At higher rates, $^7$Be becomes completely negligible but the abundance of tritium is increased so that one opens the second channel $^4{\rm He}+ ^3{\rm H}\rightarrow ^7{\rm Li}+\gamma$ so that the abundance of lithium-7 becomes too large again.

Again, it is easily concluded that in the range of parameters that allows these models to solve the lithium problem, the production of deuterium remains too high to be compatible with recent observational constraints.

\begin{figure}[htb!]
\begin{center}
 \includegraphics[width=0.9\columnwidth]{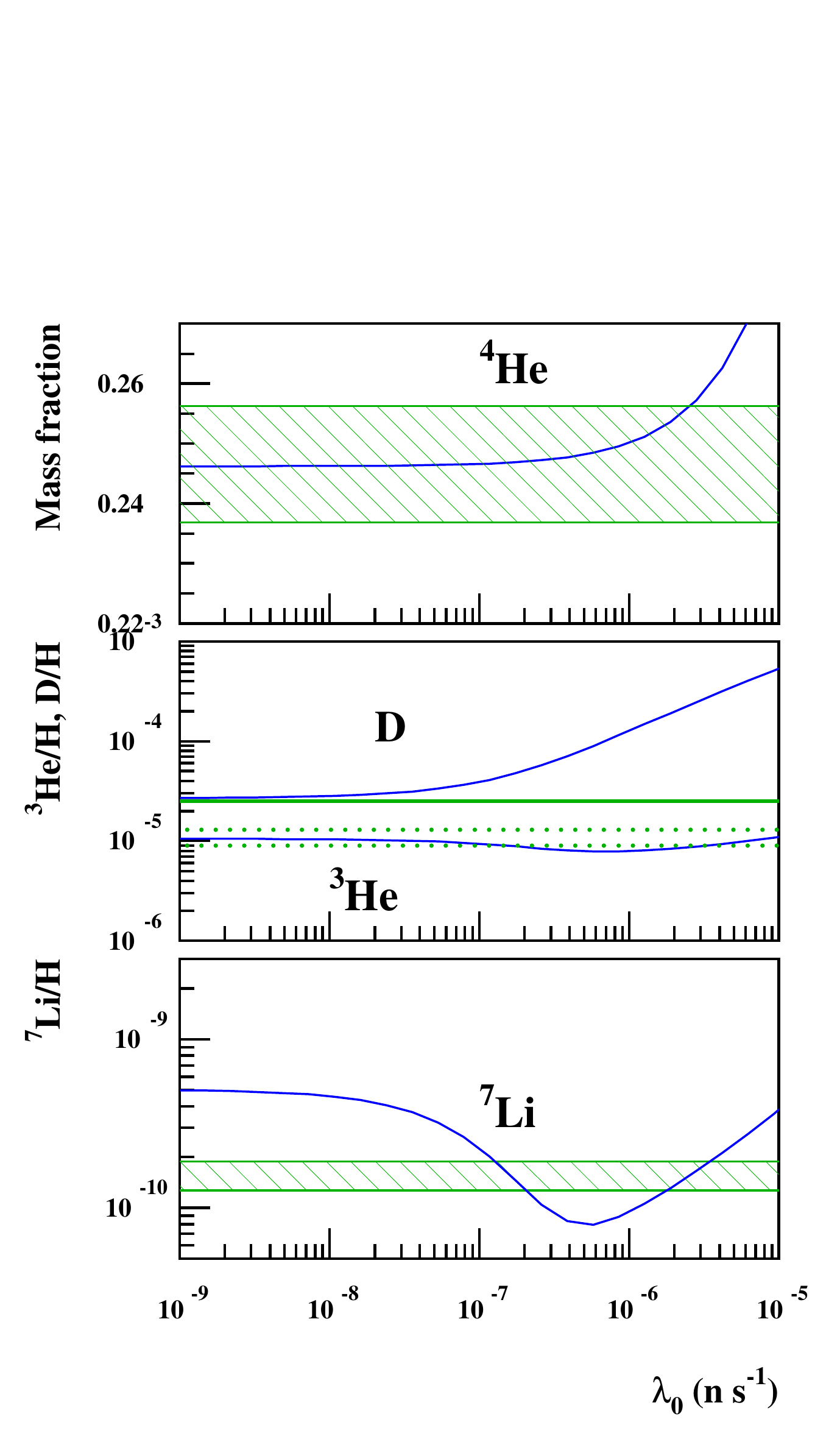}
\caption{\label{f:xxx} {\it Particle annihilation}. Abundance of  helium-4, deuterium, tritium and helium-7 as a function of the annihilation rate $\lambda_0$. Standard BBN is recovered in the limit $\lambda_0\rightarrow0$. It is easily concluded that solving the lithium-7 problem would be at the origin of deuterium problem.}
\end{center}
\end{figure}

\vskip.5cm
{\em Particle annihilation.} The only parameter of the model is the annihilation rate $\lambda_0(T/  \;GK)3$. Figure~\ref{f:xxx} depicts the dependence of the abundances of helium-4, deuterium, tritium and helium-7 as a function of this parameter assuming that $\eta$ is fixed to $\eta_{\rm CMB}$. As the annihimation rate increases, the abundance of helium-4 increases, simply because there is more neutron available. This sets an upper bound on $\lambda_0$. As already concluded in Ref.~\cite{oai:arXiv.org:1208.0443}, the neutron injection can alleviate the lithium problem. the shape of the curve is understood in exactly the same way as in the previous paragraph. While tritium is slightly affected by the neutron injection, deuterium increases and there is no possibility to reconcile both deuterium and lithium-7 simultaneously with the observations.

\begin{figure}[htb!]
\begin{center}
 \includegraphics[width=0.9\columnwidth]{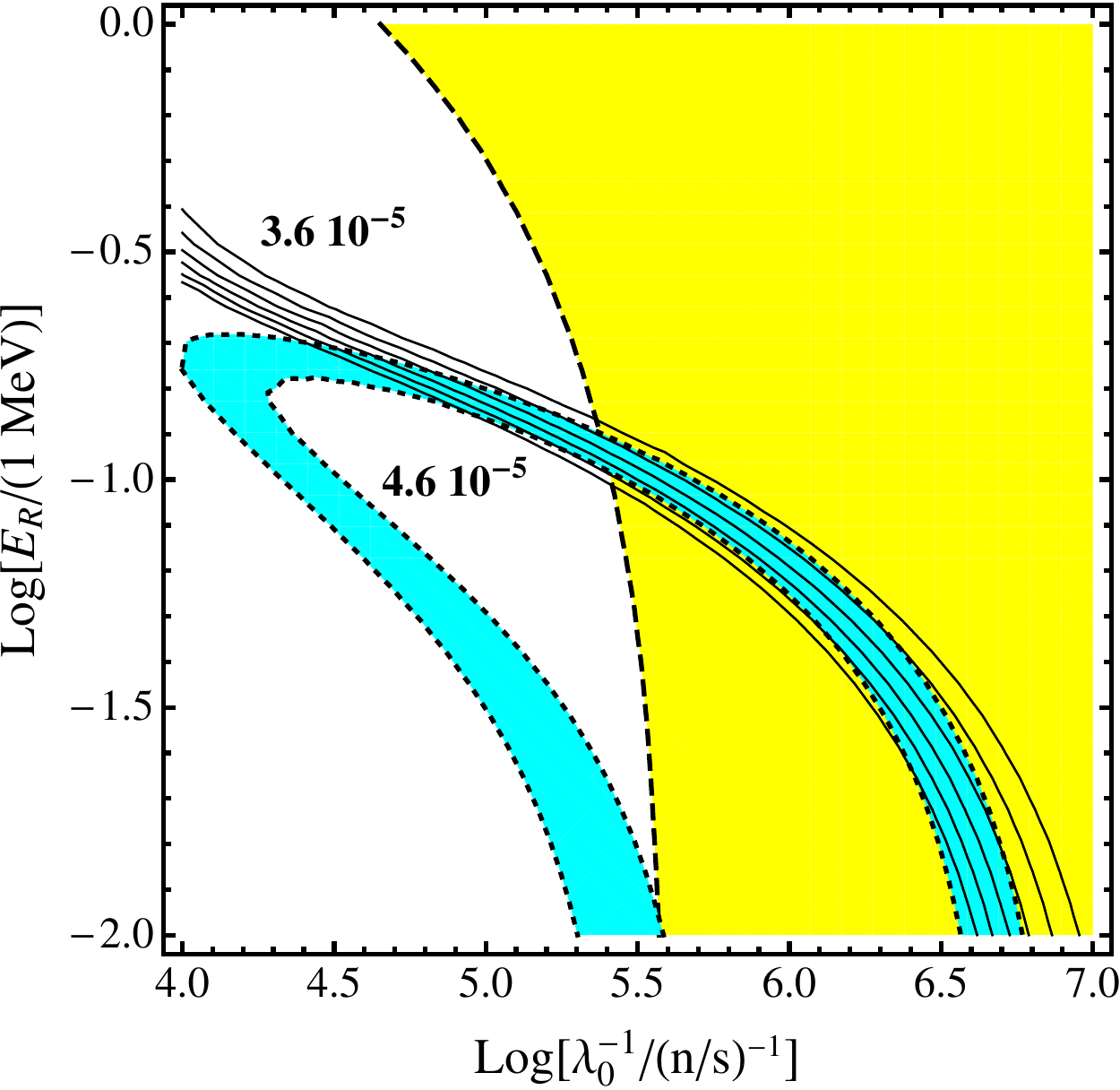}
\caption{\label{f:reson} {\it Resonnant annihilation}. Contour plot assuming $\eta=\eta_{\mathrm CMB}$ for the two parameters of the model: the resonance energy $E_R$ and the reaction rate $\lambda_0\exp(-E_R/\mathrm{k}T)$ (this corresponds to the case 5 of Ref.~\cite{oai:arXiv.org:1208.0443}). The solid dashed lines indicates the prediction of deuterium abundance D/H = \{3.6, 3.8, 4.0, 4.2, 4.4, 4.6\}$ \times10^{-5}$ from top to bottom.)}
\end{center}
\end{figure}

\vskip.5cm
{\em Resonant particle annihilation.} We scan the parameter space ($E_r,\lambda_0$) and the result is depicted on Fig.~\ref{f:reson}. The morphology of the allowed region is similar to Fig.~\ref{f:decay} obtained for particle decay. 

The morphology of the region of the parameter space leading to an agreement for both lithium-7 and helium-4 is similar to the case of the decay of a massive particle (see Fig.~\ref{f:decay}) and the existence of the two branches is interpreted in exactly the same way.

Again, the predicted abundance of deuterium is too large in these models.

\section{Conclusions}\label{sec5}

In this article we have considered four different mechanisms that allow one to modify the standard BBN framework by injecting extra neutrons during the late stages of primordial 
nucleosynthesis.  Such an injection reduces the amount of produced $^7$Be, and thus of the final $^7$Li abundances, 
since it increases its destruction due to a more efficient neutron capture. We have detailed the way to implement the oscillation of neutrons with mirror neutrons in BBN and showed that it can modify the lithium abundance only is the mirror symmetry is approximate, in the sense that $\Delta m\not=0$.

\begin{figure}[htb!]
\begin{center}
 \includegraphics[width=0.9\columnwidth]{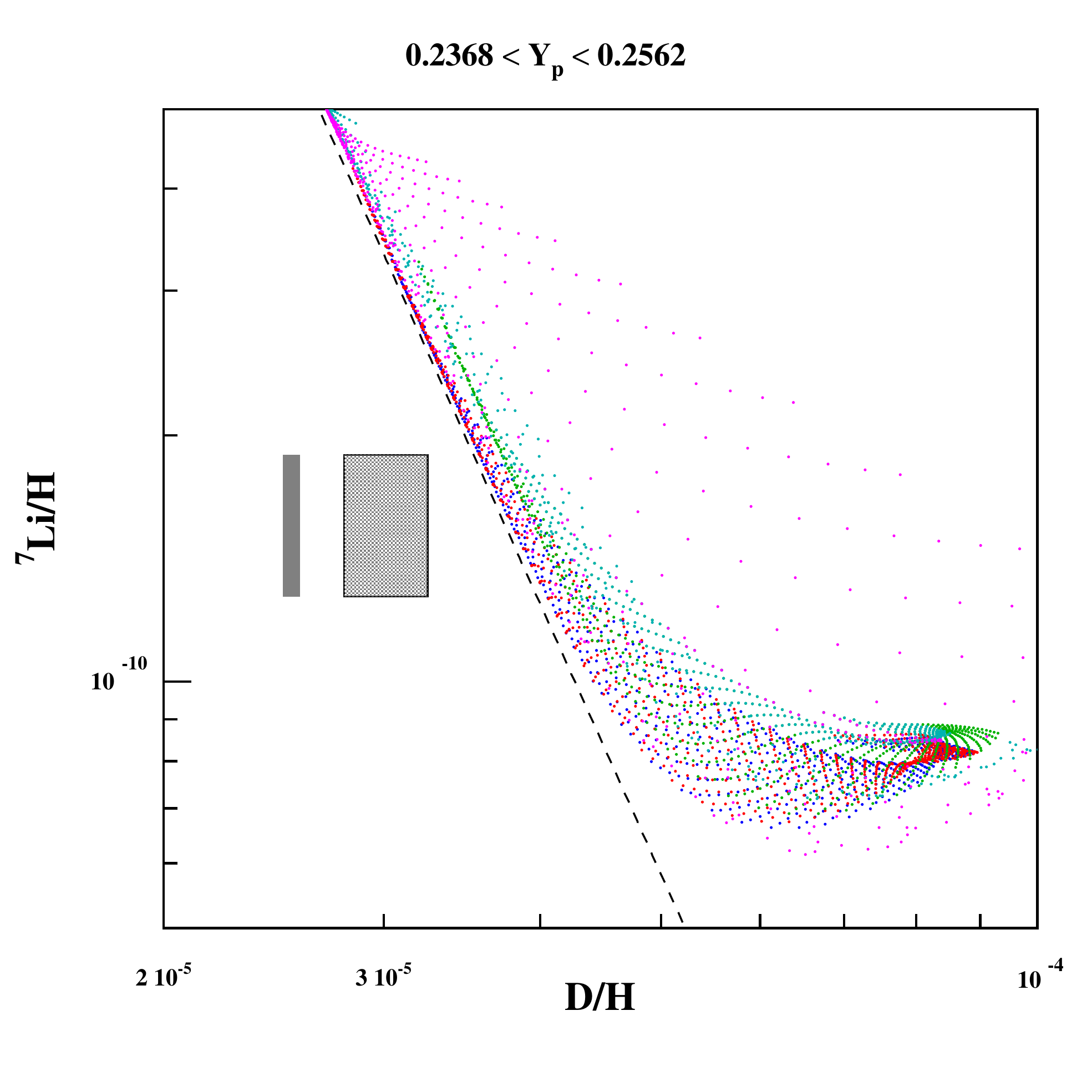}
\caption{\label{f:cor} Each dot is the prediction of a model in the space (D/H, $^7$Li/H).
The left rectangle corresponds to the D/H data of Ref.~\cite{Cooke:2013cba}  ($2.49\times10^{-5} - 2.57\times10^{-5}$) while the right rectangle corresponds to the data of Ref.~\cite{oai:arXiv.org:1203.5701} ($2.79\times10^{-5} - 3.25\times10^{-5 }$). The lithium abundance corresponds to the value of Ref.~\cite{Sbordone} ($1.27\times10^{-10} - 1.89\times10^{-10}$). This demonstrate that no model can be in agreement with both lithium-7 and deuterium. The blue, red and green dots correspond to {\it n-n' oscillation} models respectively  with $(x,\eta')=(0.2,3)$, $(x,\eta')=(0.2,1)$, $(x,\eta')=(0.5,1)$; the light blue dots correspond to {\it  resonant annihilation} models and the pink dots to {\it particle decay} models.}
\end{center}
\end{figure}

Our main conclusion is that while for all models there exists a region of the parameter space  for which both the helium-4 and lithium-7 predictions are in agreement with their current observations, assuming that $\eta$ is fixed to its CMB value, this is at the expense of a too high value of D/H, incompatible with existing observational constraints. This conclusion is summarized on Fig.~\ref{f:cor} in which each dot is the prediction of a model of one the 4 classes in the space (D/H, $^7$Li/H). It is easily concluded that all the models lies on the half-plane above the dashed line, that is
$$
\log({\rm D/H}) > -0.293 \log({}^7{\rm Li/H}) - 7.3.
$$
As a consequence, none of the models can be compatible with existing constraints on D/H (Ref.~\cite{Cooke:2013cba} or Ref.~\cite{oai:arXiv.org:1203.5701} represented by the two rectangles).

We have thus demonstrated that, given the new observational constraints on D/H, no mechanism of a neutron injection during the late stages of BBN can resolve the lithium problem. Similar conclusions for late time nucleon injection were recently reached in Ref.~\cite{kusakabe1}.

As discussed in the introduction, the solution to this problem can be from astrophysical origin or physical origin. 
In the latter case, mechanisms based on a modification of gravity (e.g. scalar-tensor theories), variation of fundamental couplings or neutron injection do not offer solutions to the lithium problem. 
Of course, one can have a combination of different mechanisms that can achieve the reduction of lithium-7 and keep deuterium unchanged ({\em e.g.} neutron injection that reduces lithium, with subsequent 
relatively soft energy injection that reduces deuterium to observable level \cite{oai:arXiv.org:1006.4172}), but such models appear to be additionally tuned. 
A partial solution to lithium problem can be achieved via the soft energy injection due to the late decay of sterile neutrinos \cite{kusakabe2}. 
Perhaps one of the most interesting remaining possibilities is the catalytic destruction of lithium via formation of the 
bound states of metastable negatively charged massive particles with nuclei, that has a potential 
of solving lithium problem without affecting deuterium \cite{cbbn}.

 It is worth emphasizing that the solution can also been of cosmological origin and lies in 
stepping away from too strict a use of the Copernican principle~\cite{regis}. While computing the abundances of the light elements during BBN, one uses the value of $\eta$ infered from CMB observation, that is a value averaged on the observable universe. The lithium spectroscopic abundances are however determined in a very local zone around our worldline (and more specifically in the Milky Way stars)
while the deuterium measurements are performed at a redshift $z\sim3$. Any large primordial downward fluctuation $\eta$, isolated in space and coincident with a position 
of the Milky Way, may just achieve the required reduction of lithium-7 locally without affecting global determination of $\eta$.

While, because of inherent doubts about the fidelity with which the Spite plateau reproduces the primordial lithium abundance, 
it is admissible to think that the cosmological lithium problem may indeed be in a category of the ``astrophysical puzzles" 
rather than be an immediate make-or-break challenge to the standard cosmological paradigm. In this latter case this problem can offer one of the rare hint of physics beyond the standard model and beyond the $\Lambda$CDM model. Our analysis shows that the recent improvement of the astrophysical data reduces the set of viable models.

\begin{acknowledgements}
MP would like to thank the IAP for the hospitality extended to him during his visit.
Research at the Perimeter Institute is supported in part by the
Government of Canada through NSERC and by the Province of Ontario through MEDT.
JPU thanks NIThEP and the University of Cape Town for hospitality during the late stages of this work.
This work made in the ILP LABEX (under reference ANR-10-LABX-63) was supported by French state funds managed by the ANR 
within the Investissements d'Avenir programme under reference ANR-11-IDEX-0004-02 and by the ANR VACOUL, ANR-10-BLAN-0510.
\end{acknowledgements}

\end{document}